\documentclass[preprint,aps,nofootinbib]{revtex4}

\usepackage{graphicx}

\usepackage{amsmath}
\usepackage{amsfonts}
\usepackage{amssymb}
\usepackage{color}

\def\be{\begin{equation}}
\def\ee{\end{equation}}
\def\bea{\begin{eqnarray}}
\def\eea{\end{eqnarray}}

\def\slashchar#1{\setbox0=\hbox{$#1$}           
   \dimen0=\wd0                                 
   \setbox1=\hbox{/} \dimen1=\wd1               
   \ifdim\dimen0>\dimen1                        
      \rlap{\hbox to \dimen0{\hfil/\hfil}}      
      #1                                        
   \else                                        
      \rlap{\hbox to \dimen1{\hfil$#1$\hfil}}   
      /                                         
   \fi}

\begin{document}

\vspace*{1cm}

\title{Double Soft Theorems and Shift Symmetry in Nonlinear Sigma Models}

\author{\vspace{0.5cm} Ian Low}
\affiliation{\vspace{0.5cm}
\mbox{High Energy Physics Division, Argonne National Laboratory, Argonne, IL 60439}\\
\mbox{Department of Physics and Astronomy, Northwestern University, Evanston, IL 60208} \\
 \vspace{0.5cm}
}

\begin{abstract}
\vspace{1cm}
We show  both the leading and subleading double soft theorems of the nonlinear sigma model follow from a shift symmetry enforcing   Adler's zero condition in the presence of an unbroken global symmetry. They do not depend on the underlying coset $G/H$ and are universal infrared behaviors of Nambu-Goldstone bosons. Although  nonlinear sigma models  contain an infinite number of interaction vertices, the double soft limit is determined entirely by a single four-point interaction, together with the existence of Adler's zeros.

\end{abstract}

\maketitle

\section{Introduction}
\label{sec:intro}

The study of soft massless particles has a long and rich history \cite{Low:1954kd,Low:1958sn,Weinberg:1964ew, Adler:1964um, Dashen:1969ez,Weinberg:1966hu}. The subject  gained renewed interest following Refs.~\cite{ArkaniHamed:2008gz, Strominger:2013jfa}, which stimulated many new analyses \cite{Casali:2014xpa,Cachazo:2015ksa,Bern:2014vva,Du:2015esa}. More recently Ref.~\cite{Cachazo:2015ksa} proposed new double soft theorems for massless scalars in a variety of quantum field theories, based on the study of scattering equations \cite{Cachazo:2014xea}. In particular, for  nonlinear sigma models (NLSM) with a $U(N)$ color (flavor) structure, they presented the following double soft theorems for a color-ordered partial amplitude
\be
\label{eq:soft1}
{M}(1,2,\cdots, n, n+1, n+2) =  \left(S^{(0)}+S^{(1)}\right) {M}(1,2,\cdots, n) + {\cal O}(\tau^{2}) \ ,
\ee
where momenta of particles $n+1$ and $n+2$ are taken soft, $p_{n+1}^\mu=\tau p^\mu$ and $p_{n+2}^\mu=\tau q^\mu$, as $\tau\to 0$. The leading and subleading soft factors are
\bea
S^{(0)}&=& \frac12\left(\frac{p_n\cdot (p_{n+1}-p_{n+2})+p_{n+1}\cdot p_{n+2}}{p_n\cdot (p_{n+1}+p_{n+2})+p_{n+1}\cdot p_{n+2}}
  + \frac{p_1\cdot (p_{n+2}-p_{n+1})+p_{n+2}\cdot p_{n+1}}{p_1\cdot (p_{n+2}+p_{n+1})+p_{n+2}\cdot p_{n+1}}\right)\ ,\\
S^{(1)}&=& \frac{p_{n+1,\mu}p_{n+2,\nu}} {p_n\cdot (p_{n+1}+p_{n+2})+p_{n+1}\cdot p_{n+2}} J_n^{\mu\nu} +  
  \frac{p_{n+2,\mu}p_{n+1,\nu}} {p_n\cdot (p_{n+2}+p_{n+1})+p_{n+2}\cdot p_{n+1}} J_1^{\mu\nu} \ ,
\eea
where $J_a^{\mu\nu}$ is the total angular momentum operator acting on the $a^{\rm th}$ scalar particle
\be
\label{eq:ang}
J_a^{\mu\nu} \equiv p_a^\mu \frac{\partial}{\partial p_{a,\nu}} - p_a^\nu \frac{\partial}{\partial p_{a,\mu}} \ .
\ee 
The leading double soft factor $S^{(0)}$ contains the famous Adler's zero \cite{Adler:1964um} when either $p^\mu\to 0$ or $q^\mu\to 0$. Expanding to the first order in $\tau$, it also reproduces the double soft limit proposed in Ref.~\cite{ArkaniHamed:2008gz},\footnote{Early studies on the double soft pion emission, using techniques of current algebra, can be found in \cite{Weinberg:1966hu,Dashen:1969ez}} which was  later studied using BCFW-like recursion relations \cite{Kampf:2013vha}. The subleading double soft factor $S^{(1)}$ was  proven with also recursion relations \cite{Du:2015esa}. 

However, it is well-known from the early work of soft pion theorems \cite{Adler:1964um,Dashen:1969ez,Weinberg:1966hu} that emission of soft pions often are uniquely determined by current algebra, i.e. commutators of vector and axial currents. Therefore, they are dictated by the Ward identities, which are statements of the symmetry in the system. Indeed, many of the recent results concentrate on relating  soft theorems to symmetries \cite{Strominger:2013jfa,Bern:2014vva}. This is the viewpoint we wish to pursue. In particular, we will see that it is possible to derive the double soft theorems for the full scattering amplitudes, i.e. the $S$-matrix elements, without recourse to color-ordered partial amplitudes.

More specifically, we consider a set of scalars $\pi^a$ transforming under  an unbroken global symmetry group $H$ and impose a set of shift symmetries to forbid a scalar mass term,
\be
\pi^a \to \pi^a +\epsilon^a +\cdots \ ,
\ee
We will see that this condition, together with the requirement of preserving the global symmetry $H$ for interaction vertices, is sufficient to prove the double soft theorems in Eq.~(\ref{eq:soft1}), without recourse to BCFW-like recursion relations. The proof is similar in spirit to the derivation of soft-gluon and soft-graviton theorems using on-shell gauge invariance at tree-level \cite{Low:1958sn,Bern:2014vva}.

This work is organized as follows. We first clarify the relation between shift symmetry and Adler's zeros in the next Section, and establish the four-point interaction satisfying both the Adler's zero condition and the unbroken global symmetry group $H$.  Derivation of the double soft theorems in NLSM for the full amplitude is presented in Section \ref{sect:derivation}. We end with the Discussions section and also comment on how to recover the double soft theorem for color-ordered partial amplitudes in Eq.~(\ref{eq:soft1}) from our results.

\section{Shift Symmetry and Adler's Zeros}
\label{sect:shift}

We start with a set of scalars $\pi^a$ furnishing a linear representation of an unbroken global symmetry group $H$, whose group generators $T^r$ satisfy the Lie algebra
\be
[T^r, T^s] = i f^{rst}T^t\ .
\ee
Under an infinitesimal action of $H$, the scalars transform as
\be
\label{eq:Htransf}
\pi(x)^a \to \pi^a(x) +i \alpha^r (T^r)_{ab} \pi^b(x)  \ ,
\ee
where $(T^r)_{ab}$ is the matrix entry of the generator $T^i$ in the particular representation under consideration. Similar to Ref.~\cite{Low:2014nga}, we  adopt a basis where $T^r$ is purely imaginary and anti-symmetric, so that all scalar fields are taken to be real. This is equivalent to writing a complex scalar in terms of its real and imaginary components.

We further assume there is a set of shift symmetries acting on $\pi^a$, 
\be
\label{eq:constshift}
\pi^a \to \pi^a +\epsilon^a +\cdots\ ,
\ee
where $\cdots$ contains higher order terms we ignore for now. Assuming the lagrangian is invariant under the shift symmetry,
\be
\label{eq:linv}
{\cal L}[\pi] \to {\cal L}[\pi] \qquad {\rm for} \qquad \pi^a \to \pi^a + \epsilon^a \ ,
\ee
a scalar mass term is forbidden and $\pi^a$'s are strictly massless.

Without specifying any details of the theory, one can derive the Ward identity corresponding to the shift symmetry in the path integral formalism \cite{Peskin:1995ev}, which gives
\bea
\label{eq:wardx}
&& \partial^\mu \langle A_\mu^a(x) \pi^{a_1}(x_{1}) \cdots \pi^{a_n}(x_{n}) \rangle \nonumber \\
&&\qquad\qquad = i \sum_r \langle \pi^{a_1}(x_{1}) \cdots\pi^{a_{r-1}}(x_{r-1}) \delta^{aa_r}\delta^{(4)}(x-x_r)\pi^{a_{r+1}}(x_{r+1})\cdots \pi^{a_n}(x_{n})\rangle \ ,
\eea
where the Noether current corresponding to the shift symmetry is
\be
A_\mu^a = \frac{\delta {\cal L}}{\delta \partial^\mu\pi^a} \ .
\ee
If we take $n=1$ in Eq.~(\ref{eq:wardx}) and Fourier-transform with respect to $\pi^{a_1}(x_1)$, the Lehmann-Symanzik-Zimmermann (LSZ) reduction formula then implies
\be
\frac{i}{p^2}\langle 0| \partial^\mu A_\mu^a(x) |\pi^{a_1}(p_1)\rangle = i f_\pi  \delta^{aa_1} e^{-ip_1\cdot x} \ ,
\ee 
leading to the famous result
\be
\label{eq:famous}
\langle 0|  A_\mu^a(x) |\pi^{a_1}(p)\rangle=i f_\pi  \delta^{aa_1}p_\mu  e^{-ip\cdot x} \ .
\ee
In other words, the Noether current $A_\mu^a$ has a non-vanishing matrix element between vacuum and the one-particle state and, therefore, can create a one-particle pole for $\pi^a$ in the correlation functions. The dimensionful parameter $f_\pi$ plays the role of the pion decay constant in low-energy QCD. For $n\ge 2$, Eq.~(\ref{eq:wardx}) and the LSZ reduction formula imply the current conservation 
\be
\label{eq:conserve}
p^\mu \langle f | \tilde{A}^a_\mu(p) | i \rangle = 0 \ ,
\ee
since the right-handed side of Eq.~(\ref{eq:wardx}) contains only $n-1$ scalar fields and has a vanishing residue when all $n$ scalar momenta are taken on-shell. As is familiar in low-energy theorems for pions, Eq.~(\ref{eq:famous}) together with the current conservation imply  the single soft limit of the scattering amplitudes of $\pi^a$'s vanishes  \cite{coleman} :
\be
\lim_{p^\mu \to 0} \langle f+\pi^a(p)| i \rangle = 0 \ ,
\ee
which is the Adler's zero condition.

The shift symmetry in Eq.~(\ref{eq:constshift}) implies the effective lagrangian must be derivatively coupled, and at lowest order it contains only the kinetic energy,
\be
{\cal L} = \frac12 \partial_\mu \pi^a\partial^\mu \pi^a + \cdots \ ,
\ee
where terms neglected are higher dimensional operators containing derivative couplings. Beyond leading order, Ref.~\cite{Low:2014nga} proposed extending $\pi^a\to \pi^a+\epsilon^a$ to second order,
\be
\label{eq:shift2}
\pi^a \to \pi^a + \epsilon^a -\frac1{3f_\pi^2} (T^r)_{ab}(T^r)_{cd}\pi^b\pi^c\epsilon^d \ ,
\ee
which form is dictated by the simple requirement of 1) invariance under the linearly realized global symmetry group $H$ and 2) it reduces to $\pi^{a_i}\to \pi^{a_i}+\epsilon^{a_i}$ when all the other scalars $\pi^a, a\neq a_i$ is turned off and set to zero, so as to fulfill the Adler's zero condition for $\pi^{a_i}$.\footnote{The numerical coefficient in the higher order term in Eq.~(\ref{eq:shift2}) is arbitrary and can be absorbed into the normalization of $f_\pi$ \cite{Low:2014nga}.}  The lagrangian invariant under the second order shift symmetry in Eq.~(\ref{eq:shift2}) is
\be
\label{eq:effL6}
{\cal L}= \frac12 \partial_\mu \pi^a\partial^\mu \pi^a - \frac1{6f_\pi^2}  (T^r)_{ab}  (T^r)_{cd}\ \partial_\mu \pi^a  \pi^b  \pi^c \partial^\mu \pi^d \ .
\ee
In fact, the full effective lagrangian for the NLSM, which is usually constructed using the formalism of CCWZ \cite{Coleman:sm,Callan:sn},  can be reproduced to all orders in $1/f_\pi$, without specifying the underlying coset $G/H$, if the following "Closure Condition" is satisfied  \cite{Low:2014nga},
\be
\label{eq:closure}
(T^i)_{ab}(T^i)_{cd} + (T^i)_{ac}(T^i)_{db}+(T^i)_{ad}(T^i)_{bc} = 0 \ .
\ee
This can be viewed as a consistency condition imposed on the low-energy effective theory constructed from the shift symmetry in the infrared. When comparing with the CCWZ formalism based on a particular coset $G/H$,  the the matrix element $(T^i)_{ab}$ should be identified with the structure constant of the broken group $G$ in the ultraviolet,
\be
\label{eq:identity}
(T^i)_{ab}=-if^{iab}={\rm Tr}(T^i[X^b,X^a])
\ee
where $T^i$ and $X^a$ are generators of $H$ and $G/H$, respectively.\footnote{With a slight abuse of notation, $(T^i)_{ab}$ in the left-hand side of Eq.~(\ref{eq:identity}) denotes the generator of $H$ in the representation under which the scalars $\pi^a$ transform, while $T^i$ in the right-hand side of the equation sits in the adjoint representation of the broken group $G$.} Then the Closure condition is nothing but the Jacobi identity of the structure constants \cite{Low:2014nga}. Eq.~(\ref{eq:closure}) is satisfied quite commonly, for examples the adjoint representation of any Lie group, as well as the fundamental representation of $SO(N)$. It is, however, not fulfilled by the fundamental representation of $SU(N)$. In this case one  need to enlarge $SU(N)$ to $SU(N)\times U(1)$ then the Closure condition is met.

At the order of $1/f_\pi^2$, it is possible to work out the form of the dimension-six operator via CCWZ to be ${\rm Tr}([X^a,X^b][X^c,X^d]) \partial\pi^a \pi^b \pi^c \partial\pi^d$, thereby giving support to the identification $(T^i)_{ab}=-if^{iab}$. However, the major distinction is the shift symmetry only requires infrared data on the group generators of the unbroken group $H$, while CCWZ requires specifying the ultraviolet data such as the broken group $G$. 

Although the full CCWZ lagrangian for the nonlinear sigma model can be obtained without knowledge of the underlying coset $G/H$, for the purpose of deriving the double soft theorems we only need the lagrangian, up to the order of $1/f_\pi^2$, in Eq.~(\ref{eq:effL6}).

\section{A derivation of the Double Soft Theorems}
\label{sect:derivation}

The double soft theorems considered in Ref.~\cite{Cachazo:2015ksa} are for NLSM with a $U(N)$ color structure and  when the two soft momenta are adjacent to each other in the color-ordered partial amplitudes. We will  avoid both assumptions by working instead with the full amplitudes, i.e. the S-matrix elements. If we define ${\cal M}^{a_1a_2\cdots a_n}(p_1,p_2,\cdots,p_n)$ to be the scattering amplitudes of $n$ scalars, it is related to the color-ordered partial amplitudes $M_\sigma(p_1,\cdots,p_n)$ by
\be
{\cal M}^{a_1a_2\cdots a_n}(p_1,p_2,\cdots,p_n) = \sum_{\sigma\in S_n/Z_n} {\rm Tr}( T^{a_{\sigma(1)}}T^{a_{\sigma(2)}}\cdots T^{a_{\sigma(n)}})
  M_\sigma(p_1,\cdots, p_n) \ ,
\ee
where the sum is over all permutations of the $n$ indices modulo cyclic permutations.  We also follow the convention that all momenta are incoming. Moreover, $M(p_1,\cdots,p_n)\equiv M_\sigma(p_1,\cdots,p_n)$ for $\sigma$ = the identity.

As was demonstrated in the previous Section,  shift symmetry imposes Adler's zeros on the amplitudes
\be
\label{eq:singlesoft}
\forall \ i \ , \quad \lim_{\tau\to 0} {\cal M}^{a_1a_2\cdots a_n}(p_1,p_2,\cdots,\tau p_i, \cdots, p_n) = 0 \ ,
\ee
while invariance under the linearly realized global symmetry $H$ requires
\be
\label{eq:Hinvariance}
\forall \ r\ , \quad \sum_{i=1}^{n} (T^r)_{a_i b}\, {\cal M}^{a_1\cdots a_{i-1} b a_{i+1}\cdots a_n}(p_1,p_2,\cdots,p_n) = 0 \ .
\ee
The above constraints can be understood  by considering ${\cal M}^{a_1\cdots a_n}$ as a rank-$n$ tensor of the unbroken group $H$. Using Eq.~(\ref{eq:Htransf}),  it transforms under the action of $H$ as
\be 
{\cal M}^{a_1\cdots a_n} \to {\cal M}^{a_1^\prime \cdots a_n^\prime} = \sum_{i=1}^n \left (1+i \alpha^r (T^r)_{a_i^\prime b}\right) {\cal M}^{a_1\cdots a_{i-1} b a_{i+1}\cdots a_n}(p_1,p_2,\cdots,p_n) \ .
\ee
Then Eq.~(\ref{eq:Hinvariance}) immediately follows by requiring ${\cal M}^{a_1\cdots a_n}={\cal M}^{a_1^\prime \cdots a_n^\prime} $. Alternatively, it can  be derived from the Ward identity associated with the $H$ invariance as well as the LSZ reduction \cite{Kampf:2013vha}.

The last ingredient we need for the derivation is the Feynman rule for the 4-point vertex in the effective lagrangian in Eq.~(\ref{eq:effL6}), written using the shorthand notation $s_{ij} \equiv (p_i+p_j)^2$,
\bea
\label{eq:4ptrule1}
&&i V^{a_1a_2a_3a_4}(p_1,p_2,p_3,p_4) =\nonumber \\
&& = \sum_{\sigma\in {\rm cycl}} \frac1{6f^2}(T^i)_{a_1a_{\sigma(2)}}(T^i)_{a_{\sigma(3)}a_{\sigma(4)}} (s_{1\sigma(4)}-s_{\sigma(2)\sigma(4)}+s_{\sigma(2)\sigma(3)} -s_{1\sigma(3)}) \ , 
\eea
where we only sum over cyclic permutations of $\{2,3,4\}$. 

Before considering the double soft limit, it is instructive to consider the single soft limit leading to the Adler's zero. The single soft amplitude receives contributions from 1) the "pole diagram" where the soft leg is attached to one of the external hard legs and 2) the gut diagram where the soft leg is attached to an internal line \cite{coleman}. When the soft momentum is taken to zero, $p^\mu=\tau p^\mu$ and $\tau\to 0$, the pole diagram could potentially develop a soft singularity, because the propagator immediately following the soft leg can now go on-shell as $\tau\to 0$,
\be
\frac{1}{(k+\tau p)^2 - m^2} = \frac{1}{2k\cdot p\, \tau  } = {\cal O}\left(\frac1{\tau}\right) \ .
\ee
In theories with a shift symmetry the massless scalar must be derivatively coupled, which means each coupling carries a positive power of momentum and could potentially cancel the soft singularity, thereby yielding a finite contribution. For Nambu-Goldstone bosons, however, no cubic couplings exist.\footnote{For massless scalars with shift symmetry the cubic coupling must contain derivative. However all kinematic invariants formed by three  light-like momenta vanish.} 
Then the Adler's zero condition implies the gut diagrams must vanish in the limit $\tau\to 0$.

\begin{figure}[t]
\includegraphics[width=3.35in, angle=0]{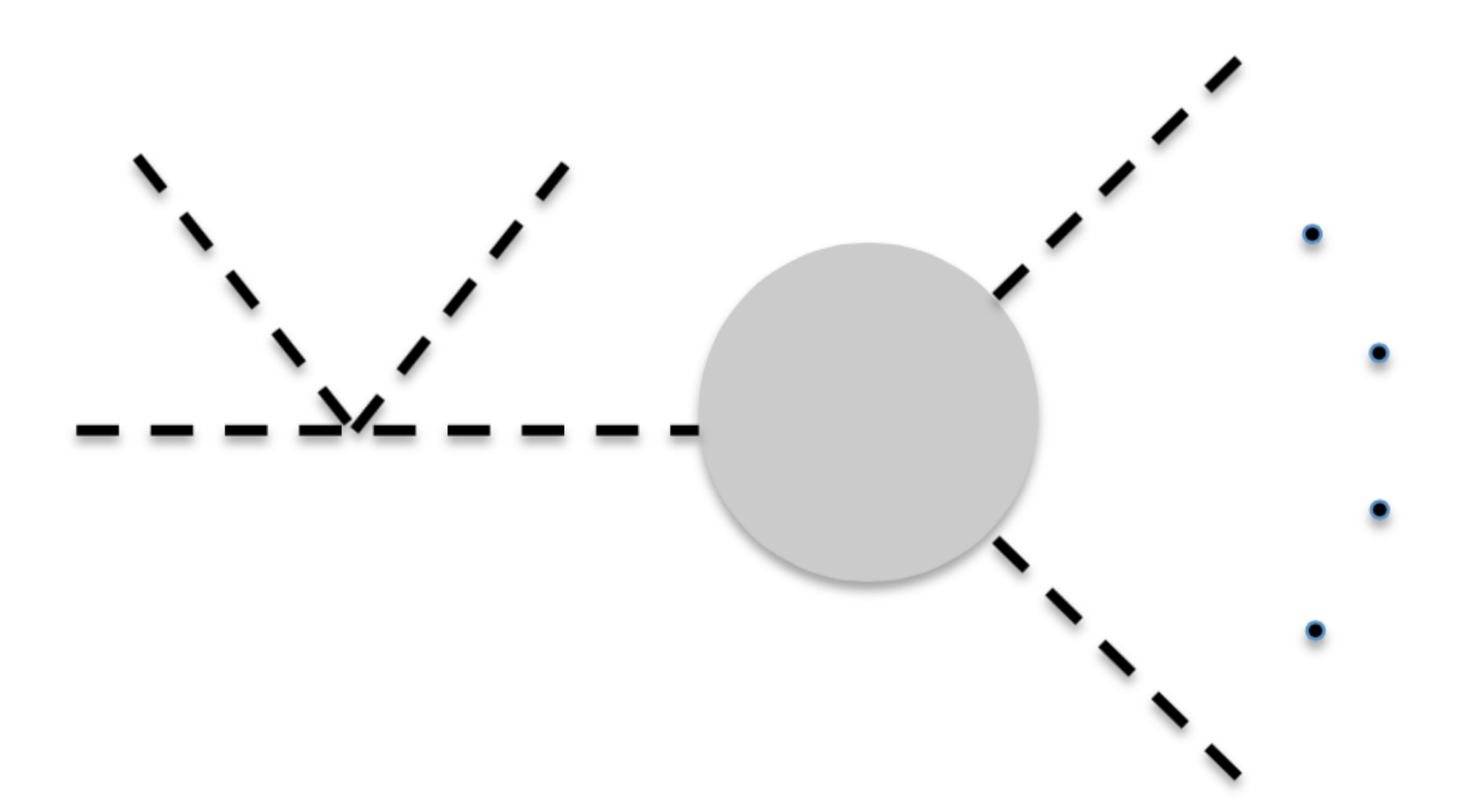}    \caption{\label{fig:1}\em An example of the pole diagram in double soft limit, where both soft legs are attached to the same external hard leg.}
\end{figure}

For double soft limit, the pole diagram now consists of diagrams where both soft legs are attached to the same external hard leg, which is shown in Fig.~\ref{fig:1}. This is the only class of diagrams where a pole in the propagator could potentially develop, while everything else belongs to the gut diagram which has no soft singularity. Thus the scattering amplitude of $n+2$ scalars can be written as
\bea
\label{eq:mn+2}
&&{\cal M}^{a_1a_2\cdots a_{n+2}}(p_1,p_2,\cdots,p_{n+2})= {N}^{a_1a_2\cdots a_{n+2}}(p_1,p_2,\cdots,p_{n+2})\nonumber \\
&&+\sum_{i=1}^n \widetilde{\cal M}^{a_ia_{n+1}a_{n+2}b}(p_i,p_{n+1},p_{n+2},q_i)\frac{1}{q_i^2}
 \widetilde{\cal M}^{a_1\cdots a_{i-1}ba_{i+1}\cdots a_{n}}(p_1,\cdots,-q_i,\cdots p_{n})\ ,
  \eea
where $N^{a_1a_2\cdots a_{n+2}}$ represents the contribution from the gut diagrams while the pole diagram factorizes into product of two semi-on-shell amplitudes, $\widetilde{\cal M}$, defined as scattering amplitudes with one of the momenta taken off-shell. Momentum conservation implies $q_i=-(p_i+p_{n+1}+p_{n+2})$ and its associated propagator in the pole diagram becomes on-shell when  $p_{n+1}$ and $p_{n+2}$ become soft simultaneously. The four-point semi-on-shell amplitude can be obtained from the Feynman rule in Eq.~(\ref{eq:4ptrule1}), after using the notation $T^{abcd}=(T^r)_{ab}(T^r)_{cd}$,
\bea
\label{eq:4ptsemi}
&& \widetilde{\cal M}^{a_ia_{n+1}a_{n+2}b}(p_i,p_{n+1},p_{n+2},q_i) = \frac{1}{3f_\pi^2}\left[T^{a_ia_{n+1}a_{n+2}b}\left(s_{(n+1)(n+2)}-s_{i(n+2)}\right)\right.\nonumber\\
&& \quad\left.+ T^{a_ia_{n+2}ba_{n+1}}\left(s_{(n+1)i}-s_{(n+1)(n+2)}\right) +T^{a_iba_{n+1}a_{n+2}}\left(s_{(n+2)i}-s_{(n+1)i}\right)\right]\ ,
\eea
We are interested in the limit $p_{n+1}^\mu\to \tau p_{n+1}^\mu$ and $p_{n+2}^\mu\to \tau p_{n+2}^\mu$ become soft as $\tau\to 0$.

Before looking at the double soft limit, however, we need to consider the single soft limit first to ensure the amplitude respects Adler's zero condition in Eq.~(\ref{eq:singlesoft}), as required by shift symmetry. By taking $p_{n+1}^\mu\to 0$ and setting the resulting amplitude to zero, we obtain
\bea
&&{N}^{a_1a_2\cdots a_{n+2}}(p_1,\cdots,p_n, 0, p_{n+2}) \nonumber\\
&&\!\!\!\!\!= \frac1{3f_\pi^2}\sum_{i=1}^n \left(T^{a_ia_{n+1}a_{n+2}b}-T^{a_iba_{n+1}a_{n+2}}\right)\widetilde{\cal M}^{a_1\cdots a_{i-1}ba_{i+1}\cdots a_{n}}(p_1,\cdots,p_i+p_{n+2},\cdots p_{n}) .
\label{eq:pn1soft}
\eea
If we further take $p_{n+2}\to 0$ in the above, 
\bea
\label{eq:N00}
&&{N}^{a_1a_2\cdots a_{n+2}}(p_1,\cdots,p_n, 0, 0) \nonumber\\
&&\!\!\!\!\!= \frac1{3f_\pi^2}\sum_{i=1}^n \left(T^{a_ia_{n+1}a_{n+2}b}-T^{a_iba_{n+1}a_{n+2}}\right){\cal M}^{a_1\cdots a_{i-1}ba_{i+1}\cdots a_{n}}(p_1,\cdots,p_i,\cdots p_{n}) \nonumber \\
&& \!\!\!\!\!= \frac1{3f_\pi^2}\sum_{i=1}^n T^{a_ia_{n+1}a_{n+2}b} {\cal M}^{a_1\cdots a_{i-1}ba_{i+1}\cdots a_{n}}(p_1,\cdots,p_i,\cdots p_{n}) \ .
\eea
where the semi-on-shell $n$-point amplitude $\widetilde{\cal M}^{a_1\cdots a_{n}}$ now becomes the on-shell amplitude ${\cal M}^{a_1\cdots a_{n}}$, since all external momenta are now on-shell as both $p_{n+1}$ and $p_{n+2}$ are taken to zero. Notice the term containing $T^{a_iba_{n+1}a_{n+2}} = (T^r)_{a_ib}(T^r)_{a_{n+1}a_{n+2}}$ vanishes due to the constraints in Eq.~(\ref{eq:Hinvariance}), which arises from the unbroken global symmetry $H$. Similarly, letting $p_{n+2}^\mu\to 0$ gives the relation
\bea
&&{N}^{a_1a_2\cdots a_{n+2}}(p_1,\cdots,p_n, p_{n+1},0) \nonumber\\
&&\!\!\!\!\!= -\frac1{3f_\pi^2}\sum_{i=1}^n \left(T^{a_ia_{n+2}ba_{n+1}}-T^{a_iba_{n+1}a_{n+2}}\right)\widetilde{\cal M}^{a_1\cdots a_{i-1}ba_{i+1}\cdots a_{n}}(p_1,\cdots,p_i+p_{n+1},\cdots p_{n}) .
\label{eq:pn2soft}
\eea
Again letting $p_{n+2}\to 0$ we recover Eq.~(\ref{eq:N00}) after using constraints from invariance under the $H$ group as well as the Closure condition in Eq.~(\ref{eq:closure}).

Next we will take both $p_{n+1}$ and $p_{n+2}$ soft simultaneously. To compare with the proposed double soft theorems in Ref.~\cite{Cachazo:2015ksa}, we will keep the propagator in $1/q_i^2$ in tact without expanding in $\tau$,
\be
\frac1{q_i^2} = \frac1{(p_i+\tau p_{n+1} + \tau p_{n+2})^2}\to \frac1{\tau(s_{i(n+1)}+s_{i(n+2)})+\tau^2 s_{(n+1)(n+2)}}\ .
\ee
On the other hand, we will only keep terms up to ${\cal O}(\tau)$ in the semi-on-shell amplitudes. 

The contribution from the gut diagram gives, after expanding in power series in $\tau$, 
\bea
&&{N}^{a_1\cdots a_{n+2}}(p_1,\cdots,p_{n+1},p_{n+2}) = {N}^{a_1\cdots a_{n+2}}(p_1,\cdots,p_n,0,0) \nonumber\\
&& \qquad+ \tau\ p_{n+1}^\mu \left. \frac{\partial}{\partial \bar{p}_{n+1}^\mu}\right|_{ \bar{p}_{n+1}^\mu=0} {N}^{a_1\cdots a_{n+2}}(p_1,\cdots, p_n,\bar{p}_{n+1},0) \nonumber \\
&&\qquad +  \tau\ p_{n+2}^\mu \left. \frac{\partial}{\partial \bar{p}_{n+2}^\mu}\right|_{ \bar{p}_{n+2}^\mu=0} {N}^{a_1\cdots a_{n+2}}(p_1,\cdots, p_n,0,\bar{p}_{n+2})\ .
\eea
After plugging in the conditions on ${N}^{a_1\cdots a_{n+2}}$ in  Eqs.~(\ref{eq:pn1soft}) and (\ref{eq:pn2soft}), which are obtained from requiring Adler's zeros in the amplitudes, we arrive at
\bea
&&{N}^{a_1\cdots a_{n+2}}(p_1,\cdots,p_{n+1},p_{n+2}) = \frac1{3f_\pi^2}\sum_{i=1}^n T^{a_ia_{n+1}a_{n+2}b} {\cal M}^{a_1\cdots b\cdots a_{n}} \nonumber\\
&&+\ \tau\ \frac1{3f_\pi^2}\sum_{i=1}^n \left(T^{a_ia_{n+1}a_{n+2}b}-T^{a_iba_{n+1}a_{n+2}}\right) p_{n+2}^\mu \frac{\partial}{\partial p_i^\mu} {\cal M}^{a_1\cdots b\cdots a_{n}} \nonumber \\
&& -\ \tau \ \frac1{3f_\pi^2}\sum_{i=1}^n \left(T^{a_ia_{n+2}ba_{n+1}}-T^{a_iba_{n+1}a_{n+2}}\right) p_{n+1}^\mu \frac{\partial}{\partial p_i^\mu} {\cal M}^{a_1\cdots b\cdots a_{n}} \ ,
\eea
where again the semi-on-shell amplitudes $\widetilde{\cal M}$ have been replaced by the $n$-point on-shell amplitudes ${\cal M}^{a_1\cdots b\cdots a_{n}}\equiv {\cal M}^{a_1\cdots b\cdots a_{n}}(p_1,\cdots, p_i, \cdots, p_n)$, as all external momenta become on-shell after the expansion in $\tau$.

Similarly the $n$-point semi-on-shell amplitude in the contribution from the pole diagram in Eq.~(\ref{eq:mn+2}) can be expanded up to ${\cal O}(\tau)$,
\be
 \widetilde{\cal M}^{a_1\cdots b\cdots a_{n}}(p_1,\cdots,p_i+p_{n+1}+p_{n+2},\cdots p_{n}) = {\cal M}^{a_1\cdots b\cdots a_{n}} +(p_{n+1} + p_{n+2})^\mu \frac{\partial}{\partial p_i^\mu} {\cal M}^{a_1\cdots b\cdots a_{n}} \ ,
\ee
while the four-point semi-on-shell amplitude can be expanded in $\tau$ explicitly using Eq.~(\ref{eq:4ptsemi}).

Putting everything together, we obtain the following double soft theorems:
\be
{\cal M}^{a_1\cdots a_{n+2}}(p_1,\cdots,p_{n+2}) = \left( {\cal S}^{(0)} + {\cal S}^{(1)}_{\rm sym}+{\cal S}^{(1)}_{\rm asym}\right) {\cal M}^{a_1\cdots b\cdots a_{n}}(p_1,p_2,\cdots,p_{n})\ ,
\ee
where
\bea
\label{eq:s0}
&& {\cal S}^{(0)} = \sum_{i=1}^{n}\frac1{2f^2} (T^r)_{a_ib}(T^r)_{a_{n+1}a_{n+2}} \frac{p_i\cdot (p_{n+2}-p_{n+1})}{p_i\cdot( p_{n+1}+p_{n+2}) + p_{n+1}\cdot p_{n+2}} \ ,\\
\label{eq:s0asym}
 && {\cal S}^{(1)}_{\rm asym} = \sum_{i=1}^{n}\frac1{2f^2} (T^r)_{a_ib}(T^r)_{a_{n+1}a_{n+2}}\frac{2 p_{n+1}^\nu p_{n+2}^\mu}{p_i\cdot( p_{n+1}+p_{n+2}) + p_{n+1}\cdot p_{n+2}} J_i^{\mu\nu}  \ ,\\
 \label{eq:s0sym}
  && {\cal S}^{(1)}_{\rm sym} =   \sum_{i=1}^{n}\frac1{2f^2} \left[  (T^r)_{a_ia_{n+1}}(T^r)_{a_{n+2}b}+(T^r)_{a_ia_{n+2}}(T^r)_{a_{n+1}b}\right]\nonumber \\
  &&\qquad\qquad \qquad\qquad \times \frac{p_{n+1}\cdot p_{n+2}}{p_i\cdot( p_{n+1}+p_{n+2}) + p_{n+1}\cdot p_{n+2}} .
 \eea
The angular momentum operator $J^{\mu\nu}_i$ for the $i^{\rm th}$ particle is defined in Eq.~(\ref{eq:ang}).

\section{Discussions}

Having derived the leading and subleading double soft theorems in NLSM, we conclude with several comments and discussions:
\begin{itemize}

\item The soft theorems for the NLSM are written entirely using infrared data: $(T^i)_{ab}$ is the generator of the unbroken global symmetry group in the IR, without reference to the broken group in the UV. This reflects the fact that Nambu-Goldstone bosons interpolate the different degenerate vacua and their interactions encode the structure of the vacua. 

\item We have only used the explicit form of the four-point interaction, together with the existence of Adler's zeros, to derive the soft theorems. This observation is fairly general and should apply to other types of theories, including gluons and gravitons. In these cases the cubic interaction will also enter, when each soft leg is connected to a different external hard leg via three-point interactions. 

\item We can uplift the soft theorems to the formalism of CCWZ by using the identification in Eq.~(\ref{eq:identity}). Then the leading soft factor ${\cal S}^{(0)}$ agrees with the well-known double soft-pion theorem \cite{Dashen:1969ez,Weinberg:1966hu,ArkaniHamed:2008gz}, which is determined by the commutator of the two soft indices $[X^{a_{n+1}},X^{a_{n+2}}]$. While the leading order soft factor ${\cal S}^{(0)}$ is anti-symmetric in $a_{n+1}$ and $a_{n+2}$, the next-to-leading soft factor  contains both an anti-symmetric component ${\cal S}^{(1)}_{\rm asym}$ and a symmetric component ${\cal S}^{(1)}_{\rm sym}$. 

\item We derived the soft theorems for the full amplitudes, which include cases when the two soft legs are adjacent to each other, or when there is a hard leg sandwiched by the two soft legs. This can be seen by applying Eq.~(\ref{eq:identity}) to express the group-theoretic factor in the soft factors in color-ordered form,
\bea
\label{eq:decompose}
(T^r)_{ab}(T^r)_{cd} &=& {\rm Tr}\left([X^{a},X^b][X^{c},X^{d}]\right) \\
&=& {\rm Tr}\left(X^{a}X^bX^{c}X^{d}\right)-{\rm Tr}\left(X^{a}X^bX^{d}X^{c}\right)\nonumber \\
&&\quad+{\rm Tr}\left(X^bX^{a}X^{d}X^{c}\right)-{\rm Tr}\left(X^aX^{a}X^{d}X^{c}\right)  ,
\eea
from which we see terms that are anti-symmetric in $a_{n+1}$ and $a_{n+2}$ arises from diagrams where the two soft legs are adjacent to each other. The leading soft factor ${\cal S}^{(0)}$ receives contribution only from this class of diagrams. Diagrams where there is a hard leg sandwiched between the soft legs are next-to-leading order in the soft expansion and only contribute to ${\cal S}^{(1)}_{\rm sym}$. There is no contribution at this order in $\tau$ from other types of configurations, consistent with the finding of Ref.~\cite{Du:2015esa} using recursion relations.

\item Using the color decomposition in Eq.~(\ref{eq:decompose}), we can also recover the result for color-ordered partial amplitudes in Eq.~(\ref{eq:soft1}). More specifically, the partial amplitude $M(1,\cdots, n+2)$ receives contributions from diagrams similar to Fig.~\ref{fig:1}, where the $(n+1)^{\rm th}$ and $(n+2)^{\rm th}$ legs are attached only to either the $1^{\rm st}$ hard leg in the color-order of $\{b(n+1)(n+2)1\}$, or the $n^{\rm th}$ hard leg, in the color-order of $\{bn(n+1)(n+2)\}$. Here the index $b$ represents the "color" of the off-shell leg carrying the momentum $q_i=-(p_{n+1}+p_{n+2}+p_i)$.  Singling out these two contributions in Eq.~(\ref{eq:s0}--\ref{eq:s0sym}) reproduces Eq.~(\ref{eq:soft1}) exactly.

\end{itemize}


Last but not least, it seems plausible that the same approach of zooming in on four-point (and three-point) couplings  would allow one to derive the other double soft theorems proposed in Ref.~\cite{Cachazo:2015ksa}. One obvious question is then the connection to other approaches for deriving the soft theorems, which involve either the recursion relation  or the scattering equations. It would be interesting to clarify this connection in the future. In addition, it would also be interesting to extend the present analysis to cases including spontaneously broken spacetime symmetry.


\begin{acknowledgments}

The author would like to thank Cliff Cheung and Yu-tin Huang for comments on the manuscript. This work is supported in part by the U.S. Department of Energy under Contracts No. DE-AC02-06CH11357 and No. DE-SC0010143. 

\end{acknowledgments}



\end{document}